\documentclass[twocolumn,prl]{revtex4}
\usepackage{amsmath}
\usepackage{color}
\usepackage{ulem}
\newcommand{\ra}{\rangle}

\begin{document}

\title{Random Walk of Bipartite Spins in a Classicalized Holographic Tensor Network}

\author{Eiji Konishi}
\email{konishi.eiji.27c@kyoto-u.jp}
\address{Graduate School of Human and Environmental Studies, Kyoto University, Kyoto 606-8501, Japan}

\date{\today}

\begin{abstract}
We consider the random walk of spin-zero bipartite spins in the classicalized holographic tensor network of the ground state of a strongly coupled two-dimensional conformal field theory.
The bipartite-spin distribution induces a metric in this network.
In the steady state of the random walk, the induced metric gives the two-dimensional anti-de Sitter (AdS$_2$) space metric.
We consider this distribution as a microscopic statistical model of the AdS$_2$ space metric.
\end{abstract}

\maketitle

The holographic tensor network (HTN)\cite{Swingle,Swingle2} represents the multiscale entanglement renormalization ansatz of the state vector of a quantum system at a quantum critical point\cite{Vidal1,Vidal2} and offers a novel interpretation of the Ryu--Takayanagi formula for entanglement entropy\cite{RT1,Tak} in the anti-de Sitter space/conformal field theory (AdS/CFT) correspondence\cite{AdSCFT1,AdSCFT2}.
Recently, in Ref.\cite{Konishi}, the author addressed the holographic interpretation of the Shannon entropy of the coherence of quantum pure states in a strongly coupled CFT in the context of AdS$_3$/CFT$_2$ correspondence.
As a result, the action of the {\it classicalized} HTN (cHTN) of the ground state of the CFT$_2$ was derived.\footnote{Here, {\it classicalization} means reducing the quantum pure ensemble of a state vector to the classical mixed ensemble of product eigenstates.}

Recall that, in analytical mechanics, action has ambiguity arising from an overall multiplicative factor.
For this reason and the sake of convenience, we drop the bit factor $b=\ln 2$ from the result of the action of the cHTN of the ground state of the CFT$_2$ in Ref.\cite{Konishi}:
\begin{equation}
I_{\rm bulk}=-\hbar A_{\rm TN}\;,\label{eq:action}
\end{equation}
where $A_{\rm TN}$ is the discretized area of the cHTN.
Here, the Dirac constant $\hbar=h/2\pi$ is the quantum of angular momentum.
This action therefore indicates that the cHTN is a web in the classical mixed ensemble of product-state networks of {\it spin} eigenstates.
A key consequence from Eq.(\ref{eq:action}) is that we obtain a formula for energy per pixel $R_{\rm AdS}^2$ of the HTN:
\begin{eqnarray}
\varepsilon\equiv R_{\rm AdS}^2{\cal T}=c\hbar(8\pi \ell_P)^{-1}\;,\label{eq:energy1}
\end{eqnarray}
where ${\cal T}$ is the tension of the AdS$_2$ space, $R_{\rm AdS}$ is the curvature radius of the AdS$_3$ space-time, and $\ell_P$ is the Planck length in three-dimensional space-time.

The aims of this article are three-fold:
(i) to derive the time-evolution equation of the random walk (i.e., the discrete diffusion)\cite{PhysRep} of {\it spin-zero} bipartite spins in a gauge-fixed spin-zero cHTN of the ground state of a strongly coupled CFT$_2$ based on the energy formula (\ref{eq:energy1});
(ii) to solve this equation for the steady state; and
(iii) to calculate the metric in the cHTN induced by the bipartite-spin distribution in the steady state.

With regards to the first aim, we consider the binary HTN at a given time.
There are two distinct spatial directions in the HTN.
The first is the radial and renormalization group (RG) direction $n$,
and the second is the horizontal direction, denoted by $j_n$ or $j$, parallel to the boundary.
In our model, each site $v=(j_n,n)$ as a disentangler in the HTN contains {\it four} spin qubits to be classicalized: {\it two} spin qubits after the disentanglement
(called the {\it earlier spin qubits})
and {\it two} spin qubits in a Bell state before the disentanglement
(called the {\it later spin qubits}).
Here, we regard the RG in the HTN as an inverse time evolution with respect to its own `{\it time}' $u=-n$\cite{Tak}.
After classicalization of the HTN, we model a gauge-fixed spin-zero cHTN at a given time as a spatial and {\it undirected} graph $\Gamma$ with edges, where the horizontal edges are {\it bivalent} and between the earlier spin bits in a site and the later spin bits in a horizontally adjacent site.
In each RG step of the cHTN, horizontally adjacent sets of earlier spin bits are staggered between the spin-zero bipartite-spin states $|\uparrow\downarrow\ra$ and $|\downarrow\uparrow\ra$.

Consider a distribution function $f_\alpha(v,t_n)$ of the site (i.e., vertex) $v$ in $\Gamma$ and the proper time $t_n$ indexed by a spin-zero bipartite-spin state $\alpha=\uparrow\downarrow$, $\downarrow\uparrow$.
To write out the time-evolution equation of the bipartite-spin random walk, we define the graph Laplacian $\Delta_\Gamma$ of $\Gamma$.
Letting $e_\alpha$ be an edge of the graph $\Gamma$ and be {\it fictitiously directed}, we define two linear operators
\begin{eqnarray}
Q_v^tf_\alpha(v)&\equiv&\sum_{e_\alpha \in E_\alpha^+(v)}(f,e_\alpha)-\sum_{e_\alpha \in E_\alpha^-(v)}(f,e_\alpha)\;,\\
Q_e(f,e_\alpha)&\equiv&\sum_{v\in V_\alpha^+(e_\alpha)}f_\alpha(v)-\sum_{v\in V_\alpha^-(e_\alpha)}f_\alpha(v)
\end{eqnarray}
for the network part of the distribution function $f=f_\alpha(v)$.
Here, we introduce two kinds of sets.
The first is $E_\alpha^-(v)$ (resp., $E_\alpha^+(v)$) comprising the outgoing (resp., incoming) edges for site $v$.
Each of these edges represents a `{\it temporal}' transition between a bipartite-spin state $\alpha$ at site $v$ and the same bipartite-spin state $\alpha$ at a radially/horizontally adjacent site.
The second kind is one-element sets of the starting site with the bipartite-spin state $\alpha$, $V_\alpha^-(e_\alpha)$ and the ending site with the bipartite-spin state $\alpha$, $V_\alpha^+(e_\alpha)$ for an edge $e_\alpha$.
The graph Laplacian $\Delta_\Gamma$ is defined for each RG step $n$ as follows\cite{Laplace1}:
\begin{equation}
\Delta_\Gamma f_\alpha(v)\equiv \frac{1}{n_\alpha(v)}Q_e\bigl(Q_v^tf_\alpha(v)\bigr)\bigr|_{n{\rm \mathchar`-th\ grains}}\;,\label{eq:Horizon}
\end{equation}
where $n_\alpha(v)$ is the number of edges $e_\alpha$ leaving site $v$.

Next, we explicitly write out the action of the graph Laplacian $\Delta_\Gamma$ on the network part of the distribution function $f_\alpha(v)$, which is based on the ansatz
\begin{equation}
f_\alpha(j_n,n)=\gamma_\alpha(j_n)\eta(n)\;.\label{eq:ansatz}
\end{equation}
For RG step $n$, the action of the radial part of the graph Laplacian $\Delta_\Gamma$ is proportional to
\begin{eqnarray}
-2\eta(n+1)+6\eta(n)-4\eta(n-1)\label{eq:Swingle0}
\end{eqnarray}
since we consider the binary HTN.
For horizontal positions $j_n$, the action of the horizontal part of the graph Laplacian $\Delta_\Gamma$ is alternately (with respect to $j_n$) proportional to the pair
\begin{align}
-\gamma_{\uparrow\downarrow}(j_n+1)&+4\gamma_{\uparrow\downarrow}(j_n)-\gamma_{\uparrow\downarrow}(j_n-1)\;,\label{eq:pro1}\\
-2\gamma_{\downarrow\uparrow} (j_n+1)&+2\gamma_{\downarrow\uparrow} (j_n)-2\gamma_{\downarrow\uparrow}(j_n-1)\label{eq:pro2}
\end{align}
or the pair
\begin{align}
-2\gamma_{\uparrow\downarrow }(j_n+1)&+2\gamma_{\uparrow\downarrow}(j_n)-2\gamma_{\uparrow\downarrow}(j_n-1)\;,\label{eq:pro3}\\
-\gamma_{\downarrow\uparrow}(j_n+1)&+4\gamma_{\downarrow\uparrow}(j_n)-\gamma_{\downarrow\uparrow}(j_n-1)\;.\label{eq:pro4}
\end{align}

Now, we derive the time-evolution equation of the bipartite-spin random walk in the graph $\Gamma$.
From Eq.(\ref{eq:energy1}), we obtain the Malgolus-Levitin time $t_{\rm ML}$\cite{ML} for the {\it `temporal'} gates of bipartite-spin states in the HTN:
\begin{equation}
t_{\rm ML}=\frac{h}{4\cdot 2\varepsilon}\;.\label{eq:ML}
\end{equation}
In the horizontal random walk of spin-zero bipartite spins, we identify the elapse of half of $t_{\rm ML}$ in proper time $t_n$ with the execution of a disentangler/entangler operation in RG step $n$.
From this identification, the Markovian stochastic master equation of the bipartite-spin random walk is formulized as follows\cite{Laplace2}:
\begin{equation}
\frac{\delta_{t_n} f_\alpha(v,t_n)}{t_P}=-\frac{2}{t_{\rm ML}}\Delta_\Gamma f_\alpha(v,t_n)\;.
\end{equation}
Here, $\delta_{t_n}$ is the dimensionless proper-time-difference operation by the Planck time $t_P=\ell_P/c$.
Using Eqs.(\ref{eq:energy1}) and (\ref{eq:ML}), we can rewrite this equation simply as
\begin{equation}
\delta_{t_n} f_\alpha(v,t_n)=-\frac{1}{\pi^2}\Delta_\Gamma f_\alpha (v,t_n)\:.\label{eq:WE}
\end{equation}
In particular, for the steady state, Eq.(\ref{eq:WE}) becomes
\begin{equation}
\Delta_\Gamma f_\alpha(v)=0\;.\label{eq:equi}
\end{equation}

Using Eq.(\ref{eq:ansatz}), we solve the equation of the bipartite-spin random walk for the steady state.
For RG step $n$ in $\Gamma$, the non-trivial solution of the radial part of Eq.(\ref{eq:equi}) is obtained from Eq.(\ref{eq:Swingle0}) as follows\cite{Swingle}:
\begin{equation}
\eta(n+1)=2\eta(n)\;.\label{eq:Swingle2}
\end{equation}
The solution of the horizontal part of Eq.(\ref{eq:equi}) is obtained from Eqs.(\ref{eq:pro1})--(\ref{eq:pro4}) as
\begin{eqnarray}
\gamma_{\uparrow\downarrow}(j_n+1)&=&2^{\pm 1}\gamma_{\uparrow\downarrow}(j_n)\;,\label{eq:pro5}\\
\gamma_{\downarrow\uparrow}(j_n+1)&=&2^{\mp 1}\gamma_{\downarrow\uparrow}(j_n)\;,\label{eq:pro6}
\end{eqnarray}
where the double signs on the right-hand sides are in the same order.

Finally, to meet our third aim, we calculate the metric in the cHTN, induced by the bipartite-spin distribution function $f_\alpha(v)$ in the steady state,
\begin{equation}
g_{ab}(v)=\frac{1}{2}\sum_{\alpha}\frac{\delta \log_2f_\alpha (v)}{\delta v_a}\frac{\delta \log_2f_\alpha(v)}{\delta v_b}\;,\label{eq:Fisher}
\end{equation}
where the factor $1/2$ reflects the radially local $Z_2$ invariance with respect to the bipartite-spin state $\alpha=\uparrow\downarrow$, $\downarrow\uparrow$, and $\delta\log_2 f_\alpha(v)$ represents the execution of a gate.
Using Eqs.(\ref{eq:Swingle2})--(\ref{eq:pro6}), the induced metric $g_{ab}$ is
\begin{equation}
g_{jj}=2^{-2n}\;,\ \ g_{nn}=1\;,\ \ g_{jn}=0\;,\label{eq:FR}
\end{equation}
where $\delta v_j=2^n$ and $\delta v_n=1$ hold in the layer of the $n$-th RG step in the cHTN.
After the transformations $x=j\epsilon$ and $r=2^n\epsilon$ for the new lattice constant $\epsilon$, the induced metric of Eq.(\ref{eq:FR}) is equivalent to the Poincar${\acute{{\rm e}}}$ metric of the AdS$_2$ space with unit curvature radius:
\begin{equation}
g_{xx}=g_{rr}=r^{-2}\;,\ \ g_{xr}=0\;.
\end{equation}
From this result, the bipartite-spin distribution $f_\alpha(v)$ in the steady state gives a microscopic statistical model behind the AdS$_2$ space metric.
Namely, this bipartite-spin distribution in the cHTN is reduced to the AdS$_2$ space, and classical stochasticity of the cHTN is suppressed when all of the bipartite-spin states $\alpha=\uparrow\downarrow$, $\downarrow\uparrow$ in the cHTN are averaged out to take a macroscopic view of the cHTN.
Note that $d$ spin-zero bipartite-spin states $\alpha_1$, $\alpha_2$, $\ldots$, $\alpha_d$ per site in the cHTN are required for the microscopic statistical model of the AdS$_{d+1}$ space for $d\in {\boldsymbol N}_{\ge 2}$.

\end{document}